\documentclass[aps,prl,showpacs,raggedbottom,
amssymb,twocolumn,groupedaddress]{revtex4}
\usepackage{graphicx}
\usepackage{amsmath}
\usepackage{amsfonts}
\usepackage{amssymb}
\usepackage{epsfig,libertine}
\usepackage[libertine]{newtxmath}
\usepackage{color}
\usepackage{dsfont, hyperref, xcolor}
\usepackage{comment}
\usepackage{enumerate}
\newcommand{\Tr}[1]{\text{Tr}\left\{#1\right\}}

\newcommand{\bra}[1]{\langle#1\vert}
\newcommand{\ket}[1]{\vert#1\rangle}

\begin{document}

\title{Quantum correlations and ergotropy}

\author{Gianluca~Francica}
\email{gianluca.francica@gmail.com}
\address{Dipartimento di Fisica e Astronomia ``G. Galilei'', Universit\`{a} degli Studi di Padova, via Marzolo 8, 35131 Padova, Italy}

\date{\today}

\begin{abstract}
Understanding the role of classical and quantum correlations in work extraction is a problem of fundamental importance in thermodynamics. We approach this problem by considering that, in closed quantum systems, the maximum cyclic work extractable is equal to the ergotropy. Thus, we aim to identify and investigate the contributions to the ergotropy coming from different kinds of initial correlations (total, classical, discord and entanglement correlations). By doing so, we have introduced and studied quantifiers of correlations which are based on ergotropy.
In particular, our results suggest that only discord correlations always give a positive contribution to work extraction, while total, classical and entanglement correlations can reduce the work extraction.
\end{abstract}

\maketitle
{\it Introduction.} Quantum correlations like entanglement~\cite{einstein35,schrodinger35,bohr35} are possible thanks to the tensor product structure of the Hilbert space of a multipartite system and the superposition principle. 
These correlations cannot be generated from an uncorrelated state by using only local quantum operations and classical communication (see, e.g., Ref.~\cite{horodecki09} for a review). However, there exist quantum correlations, quantified by the quantum discord~\cite{henderson01,ollivier01}, which are due to quantum physical effects but do not necessarily involve entanglement.
In general, quantum correlations have received much attention in studies involving thermodynamics (see, e.g., Ref.~\cite{book19}).
Concerning the work extraction in closed quantum systems, the ergotropy~\cite{allahverdyan04}, defined as the maximum work extractable by using a cyclic unitary transformation, is a key quantity of fundamental importance. 
From this point of view, similarly to the work deficit~\cite{oppenheim02}, the difference between the global and total local 
ergotropy is called ergotropic gap~\cite{perarnau-llobet15,mukherjee16}, which is shown to be related to the presence of entanglement~\cite{alimuddin19}. Moreover, the transformation of equal energetic states under the unital operation has also been studied~\cite{alimuddin20}.
Furthermore, by considering an ancilla-assisted protocol, the quantum correlations are related to a possible increase of the extracted work (daemonic ergotropy), acquiring a so-called daemonic gain~\cite{francica17}.
Regarding quantum coherence, recently, the role of the initial quantum coherence in the energy basis has also been investigated~\cite{francica20,francica22}. 
Here, we adopt the same perspective of Ref.~\cite{francica20}, where the initial quantum coherence contribution to the ergotropy is identified, 
and we aim to clarify the role of initial correlations as a resource for work extraction. 
For simplicity, we will focus our discussion on bipartite systems, but our results are easily applicable for multipartite systems.

{\it Preliminaries.} Given a quantum system of two parties $A$ and $B$, with Hilbert space $\mathcal H=\mathcal H_A\otimes \mathcal H_B$, we are interested in the correlations 
of an initial state $\rho$. Let us distinguish the correlations by following Ref.~\cite{modi10}. We start by considering a state without correlations, which is a product state $\pi=\pi_A \otimes \pi_B$, where $\pi_A$ ($\pi_B$) is the reduced state of the subsystem $A$ ($B$). By considering the set $\mathcal P$ of all product states, the total correlations are quantified by $T(\rho)=\min_{\pi\in\mathcal P} S(\rho||\pi)$, where we have defined the quantum relative entropy $S(\rho||\eta)=\Tr{\rho(\ln \rho-\ln \eta)}$, which is non-negative and $S(\rho||\eta)=0$ if and only if $\rho=\eta$. We have that $T(\rho)=S(\rho||\pi_\rho)$, where $\pi_\rho=\rho_A\otimes \rho_B$, then it is equal to the mutual information  $T(\rho)=S(\rho_A)+S(\rho_B)-S(\rho)$, where we have defined the von Neumann entropy $S(\rho)=-\Tr{\rho \ln \rho}$.
Conversely, a state which is a mixture of locally distinguishable states is a classical state $\chi=\sum_{i,j} p_{ij} P^A_i\otimes P^B_j$, where $p_{ij}$ is a joint probability distribution and $P^A_i$ and $P^B_j$ form a complete set of orthogonal projectors of rank one. 
The correlations of these states are identified as classical correlations, which are $C(\chi)=T(\chi)$. By considering the set $\mathcal C$ of all the classical states, the amount of quantum correlations can be quantified by the relative entropy of discord $D(\rho)=\min_{\chi \in \mathcal C} S(\rho||\chi)=S(\rho||\chi_\rho)$, where $\chi_\rho$ is the closest classical state.
Thus, the mutual information can be expressed as $T(\rho)=D(\rho)+C(\chi_\rho)-L(\rho)$, which defines the quantity $L(\rho)$. In particular, by substituting in the equation all the definitions given, it is easy to see that $L(\rho)=S(\pi_\rho||\pi_{\chi_\rho})$~\footnote{Since $\chi_\rho$ is obtained by applying a dephasing map to $\rho$, as we will notice later, we get $\Tr{\rho \ln \chi_\rho} = \Tr{\chi_\rho \ln \chi_\rho}$, which implies $D(\rho)=S(\chi_\rho)-S(\rho)$, and similarly $L(\rho)=S(\pi_{\chi_\rho})-S(\pi_\rho)=S(\pi_\rho||\pi_{\chi_\rho})$.}. 
Finally, a state which can be prepared using only local quantum operations and classical communication is a separable state $\sigma=\sum_i p_i \pi^{(i)}_A\otimes \pi^{(i)}_B$, where $p_i$ is a probability and $\pi^{(i)}_A$ and $\pi^{(i)}_B$ are local states of the subsystems, and we indicate with $\mathcal S$ the set of all separable states.  We note that $\mathcal P \subset \mathcal C \subset \mathcal S$ and while $\mathcal S$ is convex, $\mathcal P $ and $\mathcal C$ are not. If $\rho$ is not separable then it is entangled and the entanglement can be quantified by the relative entropy of entanglement $E_{re}(\rho)=\min_{\sigma \in \mathcal S} S(\rho||\sigma)=S(\rho||\sigma_\rho)$, where $\sigma_\rho$ is the closest separable state. In particular, separable states can possess nonclassical features and $D(\sigma_\rho)$ is called quantum dissonance.
About the discord, it is shown that the closest classical state can be expressed as $\chi_\rho = \sum_{i,j} P^A_i\otimes P^B_j \rho P^A_i\otimes P^B_j$, then $D(\rho)= \min_{\{P^A_i, P^B_j\}} S(\chi_\rho) - S(\rho)$, since $\Tr{\rho \ln \chi_\rho} = \Tr{\chi_\rho \ln \chi_\rho}$.
In order to avoid the minimization procedure, the measurement induced disturbance has been introduced as $D'(\rho)=T(\rho)-T(\chi'_\rho)=S(\chi'_\rho)-S(\rho)$ where $\chi'_\rho$ is achieved by considering projectors corresponding to the marginal basis of the eigenstates of the reduced states $\rho_A$ and $\rho_B$, and of course $\pi_{\chi'_\rho}=\pi_\rho$. Thus, we get $T(\rho)=D'(\rho)+C(\chi'_\rho)$ and, if  $L(\rho)=0$, $D'(\rho)=D(\rho)$. 
For a pure state $\rho=\ket{\psi}\bra{\psi}$, if we consider the Schmidt decomposition $\ket{\psi}=\sum_i \sqrt{p_i} \ket{i_A}\otimes\ket{i_B}$, 
where $\sqrt{p_i}$ are called Schmidt coefficients,
the closest classical state is $\chi_\rho = \sum_i p_i \ket{i_A}\bra{i_A}\otimes\ket{i_B}\bra{i_B}$ and $D(\rho)=S(\chi_\rho)=-\sum_i p_i\ln p_i$. 
In this case, we get also $\sigma_\rho=\chi_\rho$ and $E_{re}(\rho)=D(\rho)$, then entanglement is the only quantum correlation. However, things are different for a pure multipartite state, where the dissonance can be non-zero.

Following Ref.~\cite{allahverdyan04}, we are now interested in extracting work from the quantum system in the initial state $\rho$ by using a cyclic unitary transformation, i.e., by considering a unitary time evolution operator $U$ generated by a time-dependent Hamiltonian which at the final time is equal to the initial Hamiltonian $H$.
Since the system is thermally isolated, the average work $W$ extracted from the system is equal to minus the change of average energy, i.e., $W=E(\rho)-E(U \rho U^\dagger)$, where the average energy is defined as $E(\rho)=\Tr{H\rho}$. The maximum of $W$ over the set of all the unitary operators $U$ is called ergotropy, $\mathcal E(\rho)$. After ordering the labels of eigenstates of $H$ and of $\rho$ such that $H=\sum_k \epsilon_k \ket{\epsilon_k} \bra{\epsilon_k}$, with $\epsilon_k\leq \epsilon_{k+1}$, and $\rho=\sum_k r_k \ket{r_k}\bra{r_k}$, with $r_k\geq r_{k+1}$, the ergotropy reads $\mathcal E(\rho)= E(\rho) - E(P_\rho)$, where $P_\rho$ is the passive state $P_\rho = \sum_k r_k \ket{\epsilon_k}\bra{\epsilon_k}$, which for instance can be obtained as $P_\rho = U \rho U^\dagger$ with $U=\sum_k \ket{\epsilon_k}\bra{r_k}$. The ergotropy is non-negative and $\mathcal E(\rho)=0$ if and only if $\rho$ is passive.
We can compare the ergotropy of two states $\rho$ and $\eta$. If the two states have the same energy, $E(\rho)=E(\eta)$, and there exists a unital channel (i.e., which maps the identity to the identity)  $\Lambda$ such that $\eta=\Lambda(\rho)$, we have $\mathcal E(\rho) \geq \mathcal E (\eta)$~\cite{allahverdyan04}.
To prove it, we recall that $\eta=\Lambda(\rho)$ if and only if $\eta \prec \rho$ 
 if and only if there exist unitary operators $U_i$ and probabilities $p_i$ such that $\eta = \sum_i p_i U_i \rho U_i^\dagger$~\cite{nielsen01}. We note that $E(P_\rho)$ is a concave function of $\rho$, then $E(P_\eta)\geq \sum_i p_i E(P_{U_i \rho U_i^\dagger})=E(P_\rho)$ since $P_\rho = P_{U \rho U^\dagger}$ for any unitary $U$, thus we get $\mathcal E(\rho)- \mathcal E (\eta) = E(P_\eta)-E(P_\rho)\geq 0$.
In general, for two states $\rho$ and $\eta$ having the same energy, it is easy to show the identity
\begin{equation}\label{identity}
\beta(\mathcal E (\rho) - \mathcal E(\eta)) = S(\eta)-S(\rho)-S(P_\rho||\rho_\beta)+S(P_\eta||\rho_\beta)\,
\end{equation}
for any thermal state $\rho_\beta=e^{-\beta H}/\Tr{e^{-\beta H}}$. 

Thus, by considering a multipartite scenario, our aim is to demonstrate a precise connection between the ergotropy $\mathcal E(\rho)$ and the correlations in the initial state $\rho$. In order to do this, we proceed in a similar way as for quantifying correlations, by searching the corresponding closest states, i.e., the states with same energy of $\rho$ and without the desired property (e.g., mutual information, entanglement or discord) which maximize the probability of being not distinguishable from $\rho$, and so minimize the quantum relative entropy~\cite{vedral97}.

{\it Correlations contributions to the ergotropy.} If initially the two parties do not interact with each other, the initial Hamiltonian is $H=H_A + H_B$, where $H_A$ ($H_B$) is the Hamiltonian of the part $A$ ($B$). We mainly focus on this case, so that $E(\pi_\rho)=E(\rho)$ and $\pi_\rho$ is the closest product state with same energy.
In analogy to the mutual information, we can define the contribution of the total correlations $\delta_T(\rho)$ such that
\begin{equation}
\mathcal E (\rho) = \mathcal E(\pi_\rho) + \delta_T(\rho)\,,
\end{equation}
and $\delta_T(\pi)=0$ if $\pi$ is a product state.
By using the identity of Eq.~\eqref{identity}, we get
\begin{equation}
\beta\delta_T(\rho) = T(\rho)-S(P_\rho||\rho_\beta)+S(P_{\pi_\rho}||\rho_\beta)\,,
\end{equation}
which reduces to the expression for $\mathcal E(\rho)$ recently introduced in Ref.~\cite{touil22} for a local thermal initial state, $\pi_\rho=\rho_\beta$. Since the quantum relative entropy is non-negative, we get the bounds
\begin{equation}
T(\rho)-S(P_\rho||\rho_\beta)\leq \beta\delta_T(\rho) \leq T(\rho)+S(P_{\pi_\rho}||\rho_\beta)
\end{equation}
for any $\beta$. We note that our quantity is different from the ergotropic gap~\cite{perarnau-llobet15,mukherjee16}, defined as $\Delta_{EG}(\rho)=\mathcal E(\rho)-\mathcal E(\rho_A)-\mathcal E(\rho_B)$. In particular, we get $\Delta_{EG}(\rho)-\delta_T(\rho)=E(P_{\rho_A}\otimes P_{\rho_B})-E(P_{\pi_\rho})\geq 0$, since $P_{\rho_A}\otimes P_{\rho_B}= U_A \otimes U_B \pi_\rho U_A^\dagger \otimes U_B^\dagger$ has an energy not smaller than the one of the passive state $P_{\pi_\rho}$.

For characterizing the discord contribution we consider the set $\mathcal C_E$ of classical states $\eta$ having the same energy of $\rho$, $E(\eta)=E(\rho)$. We note that $\pi_\rho\in \mathcal C_E$, so that the set $\mathcal C_E$ is not empty. The closest classical state $\eta_\rho\in \mathcal C_E$ is such that $\min_{\eta \in \mathcal C_E} S(\rho||\eta)=S(\rho||\eta_\rho)$.  Therefore, we define the contribution of discord correlations as
\begin{equation}
\delta(\rho)=\mathcal E(\rho) - \mathcal E (\eta_\rho)\,.
\end{equation}
The total correlations and the discord contributions are related by
\begin{equation}
\delta_T(\rho) = \delta(\rho)+\delta_C(\eta_\rho) -\delta_L(\rho)\,,
\end{equation}
where, given an initial classical state $\chi$, we have defined the classical correlations contribution as $\delta_C(\chi)=\delta_T(\chi)=\mathcal E(\chi) - \mathcal E(\pi_\chi)$, and $\delta_L(\rho)=\mathcal E(\pi_\rho) - \mathcal E(\pi_{\eta_\rho})$.
We note that these definitions of $\delta_T(\rho)$ and $\delta(\rho)$ are analogous to the definition for the quantum coherence contribution to the ergotropy~\cite{francica20}, $\mathcal E_c(\rho)=\mathcal E(\rho) - \mathcal E(\Delta(\rho))$, where $\Delta$ is the dephasing map with respect to the energy basis $\ket{\epsilon_k}$, $\Delta(\rho)$ is the closest incoherent state to $\rho$ and of course $E(\Delta(\rho))=E(\rho)$.

Similarly to $\chi_\rho$, it is easy to show that $\eta_\rho= \sum_{i,j} P^A_i\otimes P^B_j \rho P^A_i\otimes P^B_j$, where the projectors are not necessarily equal to the ones of $\chi_\rho$, since in general they minimize the entropy $S(\eta_\rho)$ with the constraint $E(\eta_\rho)=E(\rho)$ and of course $S(\eta_\rho)\geq S(\chi_\rho)$. To avoid misunderstandings, from now on we will indicate with $P'^A_i$ and $P'^B_j$ the projectors of $\chi_\rho$.
We observe that $\delta(\rho)$ is invariant under local unitary transformations, i.e., $\delta(\rho')=\delta(\rho)$ with $\rho'=U_A\otimes U_B\rho U_A^\dagger\otimes U_B^\dagger$ for any unitary local operators $U_A$ and $U_B$. This follows by noting that $P_{\rho'}=P_\rho$ and $P_{\eta_{\rho'}}=P_{\eta_\rho}$ since $\eta_{\rho'}=U_A\otimes U_B \eta_\rho U_A^\dagger\otimes U_B^\dagger$.  Furthermore, $\pi_{\rho'}=U_A\otimes U_B \pi_\rho U_A^\dagger\otimes U_B^\dagger$ and thus we get $\delta_C(\rho')=\delta_C(\rho)$ and $\delta_L(\rho')=\delta_L(\rho)$, so also $\delta_T(\rho')=\delta_T(\rho)$.
Since $\eta_\rho$ is obtained by applying a unital channel to $\rho$, we get $\delta(\rho)\geq 0$, and of course if $\chi$ is classical then $\delta(\chi)=0$. Therefore, $\mathcal E(\rho) = \mathcal E (\eta_\rho) + \delta(\rho) \geq \delta(\rho)$ and the discord contribution $\delta(\rho)$ gives a non-negative lower bound of the ergotropy. By using the identity of Eq.~\eqref{identity}, we get
\begin{equation}
\beta\delta(\rho) = D(\rho)+S(\eta_\rho)-S(\chi_\rho)-S(P_\rho||\rho_\beta)+S(P_{\eta_\rho}||\rho_\beta)\,
\end{equation}
from which we have the bounds
\begin{equation}
D(\rho)-S(P_\rho||\rho_\beta)\leq \beta\delta(\rho) \leq S(\rho||\eta_\rho)+S(P_{\eta_\rho}||\rho_\beta)\,
\end{equation}
for any $\beta$. If  $P_\rho$ is completely passive, i.e., is thermal, $P_\rho=\rho_{\beta}$, we have $\delta_T(\rho)\geq T(\rho)/\beta$ and $\delta(\rho)\geq D(\rho)/\beta$. Anyway, since the eigenvalues of $\rho$ are conserved in the unitary transformation, the passive state $P_\rho$ is not necessarily a thermal state and in this case the lower bounds can be negative and useless.
An analogous result for the classical correlations is achieved by considering an initial state $\chi$ classical and the definition $\delta_C(\chi)=\delta_T(\chi)$. Instead, for $\delta_L(\rho)$ we get
\begin{equation}
\beta\delta_L(\rho) = L(\rho)+S(\pi_{\eta_\rho})-S(\pi_{\chi_\rho})-S(P_{\pi_\rho}||\rho_\beta)+S(P_{\pi_{\eta_\rho}}||\rho_\beta)\,.
\end{equation}

It is worth to observe that if the energy spectrum is non-degenerate, $\epsilon_k < \epsilon_{k+1}$, then $\delta(\rho)=0$ implies that $\rho$ is classical. On the other hand, as we will see by considering a simple example, if the energy spectrum is degenerate we can have non-zero quantum correlations but $\delta(\rho)=0$. 

{\it Proof.} If the eigenvalues of $\eta_\rho$ are $q_k$ ordered such that $q_k\geq q_{k+1}$, its passive state is $P_{\eta_\rho}=\sum_k q_k \ket{\epsilon_k}\bra{\epsilon_k}$ and we get $\delta(\rho)=\sum_k (\epsilon_{k+1}+\epsilon_k) \sum_{n=1}^k(r_n-q_n)$. Since $\eta_\rho$ is obtained by applying a unital map to $\rho$, we have that $\sum_{n=1}^k(r_n-q_n)\geq 0 $ for any $k$, thus $\delta(\rho)=0$ implies $q_k=r_k$ for any $k$ and so $P_{\eta_\rho}=P_\rho$. Then, there exists a unitary $U$ such that $\eta_\rho = U \rho U^\dagger$, which implies $S(\eta_\rho)=S(\rho)$. Therefore, $S(\rho||\eta_\rho)=S(\eta_\rho)-S(\rho)=0$ and so $\rho=\eta_\rho$.

For a pure quantum state, $\rho=\ket{\psi}\bra{\psi}$, we have that $E(\chi_\rho)=E(\rho)$, then $\eta_\rho=\chi_\rho$. In particular, $\delta(\rho)$ is intimately related to the entanglement and reads
\begin{equation}\label{pure}
\delta(\rho)= \sum_i p_i \epsilon_i - \epsilon_1\,,
\end{equation}
where $p_i$ are the square of the Schmidt coefficients ordered such that $p_i\geq p_{i+1}$.
Then, if the fundamental level $\epsilon_1$ is non-degenerate, $\delta(\rho)=0$ if and only if there is no entanglement.
We note that the expression of Eq.~\eqref{pure} differs from the daemonic gain~\cite{francica17} only in energy levels.

\begin{figure}
[h!]
\centering
\includegraphics[width=0.7\columnwidth]{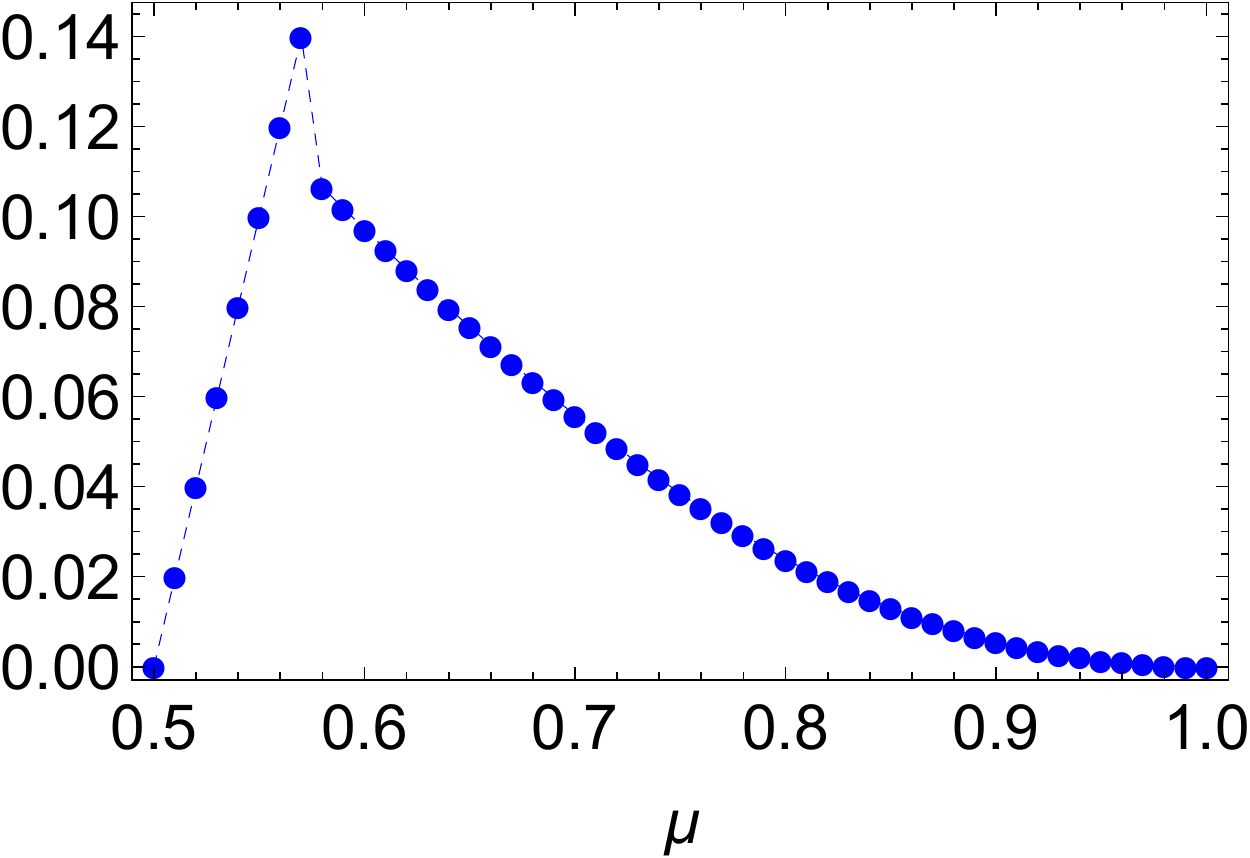}
\caption{Plot of $\delta(\rho)$ in unit of $\epsilon$ in function of $\mu$ for $R=1$. We find $\mu_c(1)\approx 0.57.$
}
\label{fig:plot01}
\end{figure}

In order to complete the discussion about the discord contribution, we consider an illustrative physical example of two qubits. We consider $H_A= R \epsilon \ket{1}\bra{1}$ and $H_B= \epsilon \ket{1}\bra{1}$, where $\{\ket{0},\ket{1}\}$ is the computational basis of a qubit, and $\epsilon,R\geq 0$. We illustrate our results by focusing on the initial state $\rho = \mu \ket{0}\bra{0}\otimes \ket{+}\bra{+} + (1-\mu)\ket{1}\bra{1}\otimes\ket{1}\bra{1}$, with $\mu \in [0,1]$ and $\ket{\pm}=(\ket{0}\pm\ket{1})/\sqrt{2}$. The projectors of the closest classical state $\chi_\rho$ are $P_1'^A=\ket{0}\bra{0}$, $P_1'^B=\ket{0}\bra{0}$ if $\mu \leq 1/2$ and  $P_1'^B=\ket{+}\bra{+}$ if $\mu>1/2$, and thus we get $D(\rho)=\min\{\mu,1-\mu\}\ln2$. The energy of the passive state is $E(P_\rho)=\min\{1,R\}\epsilon \min\{\mu,1-\mu\}$. For $\mu\leq 1/2$, since $E(\chi_\rho)=E(\rho)$, we get $\eta_\rho=\chi_\rho=\Delta(\rho)$, where the dephasing is respect to the computational basis. Then, $\delta(\rho)=\mathcal E_c(\rho)=|1-R|\epsilon \mu/2$, which is equal to zero if $R=1$ (there is degeneracy) also if the initial state is not incoherent and has quantum correlations. For $\mu > 1/2$, $\eta_\rho \neq \chi_\rho$ since $E(\chi_\rho)\neq E(\rho)$. To calculate $\eta_\rho$, we consider its projectors $P^A_1=\ket{a,\alpha}\bra{a,\alpha}$ and $P^B_1=\ket{b,\beta}\bra{b,\beta}$, with $\ket{a,\alpha}=\sqrt{a}\ket{0}+\sqrt{1-a}e^{i\alpha}\ket{1}$. The condition $E(\eta_\rho)=E(\rho)$ and $S(\eta_\rho)$ do not depend on $\alpha$, and we find $\beta = 0$, so that $E(\eta_\rho)=E(\rho)$ is equivalent to $f(a,b)=2(1-a)a R (2\mu-1)-2(1-b)b(1-\mu)+(1-2b)\sqrt{(1-b)b}\mu=0$. Thus, we determinate the set $\mathcal C_E$ by changing $b$ and finding the solutions of $f(a,b)=0$. 
Then, we find $\eta_\rho$ by minimizing $S(\eta_\rho)$ over $\mathcal C_E$. For $\mu \leq \mu_c(R)$, we find $\eta_\rho = \Delta(\rho)$, instead for $\mu>\mu_c(R)$, $\eta_\rho \neq \Delta(\rho)$. The contribution $\delta(\rho)$ is shown in Fig.~\ref{fig:plot01} and we can see that $\delta(\rho)$ is discontinuous at $\mu=\mu_c(R)$ since the set $\mathcal C_E$ is not convex.

\begin{figure}
[h!]
\centering
\includegraphics[width=0.7\columnwidth]{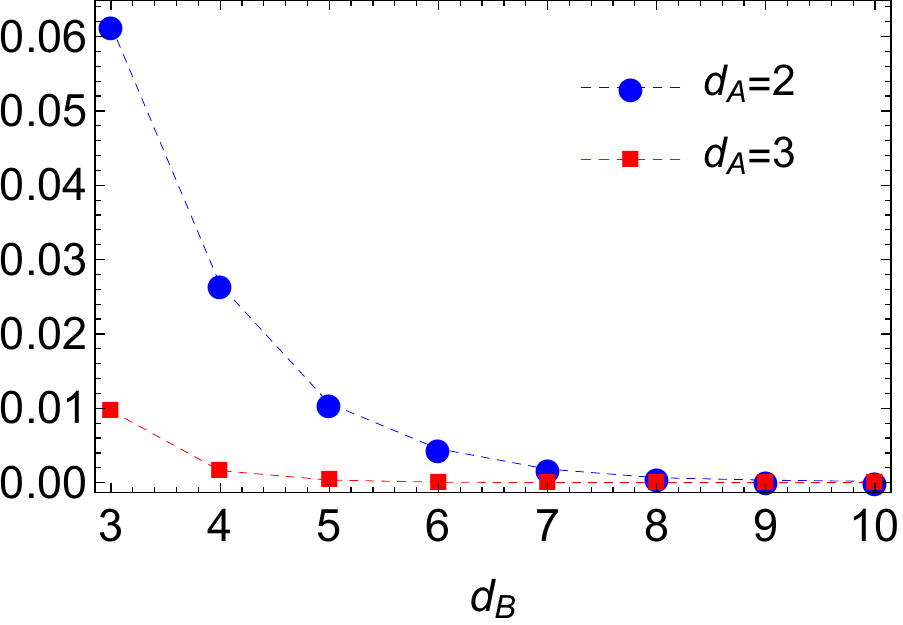}
\caption{Plot of the probability to find  $\delta_C(\chi)<0$ in function of the dimension $d_B$ for fixed $d_A$. We generate $\chi$ random and $\epsilon^A_i$, $\epsilon^B_j$ random in the interval $[0,1]$ for $n=10^5$ times and calculate the probability as the number of times to see $\delta_C(\chi)<0$ over $n$. The probability decreases exponentially with $d_B$ and tends to zero for $d_B\to\infty$.
}
\label{fig:plot02}
\end{figure}

Proceeding with the discussion of the other contributions, we note that $\delta_L(\rho)\geq 0$ since  $\pi_{\eta_\rho} = \sum_{i,j} P^A_i\otimes P^B_j \pi_\rho P^A_i\otimes P^B_j$ which is a unital channel.
On the contrary, the same is not true for the classical correlations. We consider the initial state $\chi = \sum_{i,j} p_{ij} P^A_i\otimes P^B_j$, and we start by noting that if $P_\chi=U_A\otimes U_B \chi U_A^\dagger\otimes U_B^\dagger$, the reduced states of $P_\chi$ are $U_A\chi_A U_A^\dagger$ and $U_B \chi_B U_B^\dagger$. Since $U_A$ and $U_B$ are such that the energy of $P_\chi$ is minimum, we have that $P_{\chi_A}=U_A\chi_A U_A^\dagger$ and $P_{\chi_B}=U_B \chi_B U_B^\dagger$, then $E(P_\chi)=E(P_{\chi_A}\otimes P_{\chi_B})\geq E(P_{\pi_\chi})$, from which follows $\delta_C(\chi)\leq 0$.
In general, $\delta_C(\chi)=\sum_k \epsilon_k (\tilde r_k - r_k)$, where $r_k$ and $\tilde r_k$ are the eigenvalues of $\chi$ and $\pi_\chi$ in non-increasing order, $r_k\geq r_{k+1}$ and $\tilde r_k\geq \tilde r_{k+1}$. If $d_A$ ($d_B$) is the dimension of $\mathcal H_A$ ($\mathcal H_B$), for the special case $d_A=d_B=2$, we have that $\epsilon_4-\epsilon_3=\epsilon_2-\epsilon_1$ and so $\delta_C(\chi)=(\epsilon_2-\epsilon_1)(r_1-r_4-\tilde r_1 + \tilde r_4)+(\epsilon_3-\epsilon_2)(r_1+r_2-\tilde r_1-\tilde r_2)$. We find that $r_1-r_4-\tilde r_1 + \tilde r_4\geq 0$ and $r_1+r_2-\tilde r_1-\tilde r_2 \geq 0$, then $\delta_C(\chi)\geq 0$ for any classical state $\chi$.
%
%
Therefore, we have that if $P_\chi=U_A\otimes U_B \chi U_A^\dagger\otimes U_B^\dagger$ or $\chi=\pi_\chi$ then $\delta_C(\chi)=0$, but the reverse is not true. For instance if we consider $p_{11}\geq p_{12} \geq p_{21} \geq p_{22}$ we have $\delta_C(\chi)=0$, but we can have $\chi\neq \pi_\chi$ and  $P_\chi\neq U_A\otimes U_B \chi U_A^\dagger\otimes U_B^\dagger$ (if $\epsilon^A_2<\epsilon^B_2$, where we indicate with $\epsilon^A_i$ and $\epsilon^B_j$ the eigenvalues of $H_A$ and $H_B$ in non-decreasing order). 
However, for higher dimensions $\delta_C(\chi)$ can also be negative, thus reduces the work extracted with respect to the product state. In particular, $\delta_C(\chi)$ can be negative also if $P_\chi \neq U_A \otimes U_B \chi  U_A^\dagger \otimes U_B^\dagger$. For instance, for the case $d_A=2$ and $d_B=3$, we can consider  $p_{11}=0.27$, $p_{12}=0.04$, $p_{13}=0.23$, $p_{21}=0.26$, $p_{22}=0.03$ and $p_{23}=0.17$. We get $\delta_C(\chi)=(\epsilon_2-\epsilon_1)x_1+(\epsilon_4-\epsilon_3)x_3+(\epsilon_6-\epsilon_5)x_5$ with $x_1=-0.0162$, $x_3=0.014$ and $x_5=0.0022$, and since $x_3+x_5\geq 0$ and $x_1+x_3+x_5=0$, it is easy to show that $\delta_C(\rho)<0$ if $\epsilon^A_1=\epsilon^B_1=0$, $\epsilon^A_2\leq \epsilon^B_2\leq \epsilon^B_3$ and $\epsilon^A_2+\epsilon^B_2>\epsilon^B_3$.
Anyway, a numerical analysis suggests that the classical states $\chi$ such that $\delta_C(\chi)<0$ are unlikely for large $d_A$ or $d_B$ (see Fig.~\ref{fig:plot02}).

Analogously, we can define the entanglement contribution by considering the closest separable state $\tau_\rho$ having the same energy, $E(\tau_\rho)=E(\rho)$. We define
\begin{equation}
\delta_E(\rho)=\mathcal E(\rho)- \mathcal E(\tau_\rho)\,,
\end{equation}
so that $\delta_E(\sigma)=0$ if $\sigma$ is separable. Of course also $\delta_E(\rho)$ is invariant under local unitary transformations. We note that the difference $\delta(\rho)-\delta_E(\rho)=\mathcal E(\tau_\rho)-\mathcal E(\eta_\rho)$ is related to the quantum dissonance, since it is equal to zero if $D(\tau_\rho)=0$. 
The contribution $\delta_E(\rho)$ can be expressed as $\delta_E(\rho)=\mathcal E (\pi_\rho)-\mathcal E(\tau_\rho)+\delta_T(\rho)$ and for a pure state $\rho$, we get $\delta_E(\rho)=\delta(\rho)$.
However, contrary to the discord contribution, we observe that $\delta_E(\rho)$ can be also negative for initial mixed states (only if $D(\tau_\rho)\neq 0$).
To prove it, we consider $d_A=d_B=2$, the Horodecki state $\rho=p \ket{\psi^{(\pm)}}\bra{\psi^{(\pm)}} + (1-p) \ket{00}\bra{00}$, having the closest separable state~\cite{vedral98} $\sigma_\rho = q'^2 \ket{00}\bra{00}+2p'q'\ket{\psi^{(\pm)}}\bra{\psi^{(\pm)}}+p'^2 \ket{11}\bra{11}$, where $\ket{\psi^{(\pm)}}=(\ket{01}\pm \ket{10})/\sqrt{2}$, $p'=p/2$ and $q'=1-p'$. It is easy to show that $E(\sigma_\rho)=E(\rho)$, so that $\tau_\rho=\sigma_\rho$. Thus, if we consider for instance $p=1/2$,  $\epsilon^A_1=\epsilon^B_1=0$ and $\epsilon^A_2=\epsilon^B_2=\epsilon$, we find $\delta_E(\rho)= -0.0625 \epsilon$.
Furthermore, if the inequality $S(\rho||\tau_\rho)\leq S(\tau_\rho)-S(\rho)$ is satisfied, from the identity of Eq.~\eqref{identity}, we get the lower bound $\beta \delta_E(\rho)\geq E_{re}(\rho) - S(P_\rho||\rho_\beta)$.

Finally, by considering the measurement induced disturbance, we observe that $E(\chi'_\rho)=E(\rho)$, 
 since $\pi_{\chi'_\rho}=\pi_\rho$. We define $\delta'(\rho)=\mathcal E(\rho)-\mathcal E(\chi'_\rho)$, which is non-negative since $\chi'_\rho$ is obtained by applying a unital map to $\rho$, and $\delta'(\chi)=0$ if $\chi$ is classical. By using the identity of Eq.~\eqref{identity}, we get
\begin{equation}
\beta\delta'(\rho) = D'(\rho)-S(P_\rho||\rho_\beta)+S(P_{\chi'_\rho}||\rho_\beta)
\end{equation}
and the bounds
\begin{equation}
D'(\rho)-S(P_\rho||\rho_\beta)\leq \beta\delta'(\rho) \leq D'(\rho)+S(P_{\chi'_\rho}||\rho_\beta)\,.
\end{equation}
Of course, similarly to $\delta(\rho)$ we have that if the energy spectrum is non-degenerate, then $\delta'(\rho)=0$ implies that $\rho$ is classical.
In the end, we note that $\delta_T(\rho)=\delta'(\rho)+\delta_C(\chi'_\rho)$, thus $\delta_T(\rho)$ is negative if $\delta_C(\chi'_\rho)<-\delta'(\rho)$. Then, $\delta_T(\rho)$ is non-negative if $\rho$ is a pure state, in contrast for initial mixed states, similarly to the classical correlations contribution, for $d_A=d_B=2$ we have $\delta_T(\rho)\geq 0$ and the states with a negative $\delta_T(\rho)$ are unlikely for large $d_A$ or $d_B$.

For the case of interacting parties, generalization of the correlations contributions introduced is problematic. It is enough to consider the set of the product states with the same energy of the initial state. This set can be empty. To show it, we can consider $\rho=\ket{E_{max}}\bra{E_{max}}$, where $\ket{E_{max}}$ is the state with maximum energy which we assume non-degenerate. Then, if the two parties interact with each other, $\ket{E_{max}}$ is entangled and there is not a product state having the same energy of $\rho$. Thus, for interacting parties, we cannot define the total correlations contribution by considering the closest product state with same energy. Alternatively, we can consider the closest product state $\pi_\rho$ and its energy difference with respect to the initial state to define the total correlations contribution as $\tilde \delta_T(\rho) = \mathcal E(\rho) - \mathcal E(\pi_\rho) -E(\rho)+ E(\pi_\rho)$. 
Similarly, we define the quantum correlations contribution as $\tilde \delta(\rho) = \mathcal E(\rho) - \mathcal E(\chi_\rho) -E(\rho)+ E(\chi_\rho)$ and the entanglement contribution as $\tilde \delta_E(\rho) = \mathcal E(\rho) - \mathcal E(\sigma_\rho) -E(\rho)+ E(\sigma_\rho)$.
We observe that our main finding, that only the quantum correlations always give a non-negative contribution to the ergotropy, while the total, classical and entanglement correlations can give a negative contribution, also applies to these alternative definitions. We focus on non-interacting parties, we get $\tilde \delta_T(\rho)=\delta_T(\rho)$ which can be negative, and about the entanglement contribution, for the example given above, $\tilde \delta_E(\rho)=\delta_E(\rho)<0$. About the quantum correlations contribution, we get $\tilde \delta(\rho) = E(P_{\chi_\rho}) - E(P_\rho)$ and we have $\tilde \delta(\rho) \geq 0$ as for $\delta(\rho)$, since $\chi_\rho$ is obtained by applying a unital map to $\rho$.


{\it Conclusions.} In summary, we have identified and investigated the contributions to the ergotropy coming from  different kinds of correlations by comparing the ergotropy with the one of appropriate closest states. 
By doing so, we have established an important link between the field of quantum thermodynamics and the one of quantum information theory.
Relative entropy based measure of correlations are known very well. By establishing ergotropy based quantifiers of correlations, we show how ergotropy also plays some major role to characterize quantum correlations which has less explored so far.
In particular, beyond the foundations implications, the importance of the paper is also due to the fact that the clarification of the relevance of quantum features is of central interest in order to achieve quantum thermodynamic processes with performances classically inhibited.
Interestingly, we find that only the contribution of discord correlations is always positive. Conversely, other kinds of correlations can reduce the work extraction. 
We note that the so-called free energy of the correlated state $\rho$ is not smaller than the one of the product state $\pi_\rho$ of the reduced states (see, e.g., Ref.~\cite{bera19}).
Concerning quantum correlations, the free energy of $\rho$ is not smaller than the one of $\eta_\rho$, since $S(\eta_\rho)\geq S(\rho)$, and the one of $\tau_\rho$, if $S(\tau_\rho)\geq S(\rho)$, thus the correlations contributions in terms of the free energy are expected non-negative.
We observe that things can also be different if we consider the work extraction when the system is in contact with a reservoir. If the reservoir initially is in a thermal state and uncorrelated with the system, the maximum work extractable, i.e., the ergotropy of the total system (system plus reservoir), is upper bounded by minus the change of out-of-equilibrium free energy of the system, and similar quantifiers can be defined in terms of this quantity. 
In particular, it is easy to show that the contribution of total correlations is always non-negative for non-interacting parties~\footnote{To prove it, we consider the difference of out-of-equilibrium free energy $\Delta F_{\beta}(\rho)=S(\rho||\rho_\beta)/\beta$, and by noting that, for non-interacting parties, $S(\rho||\rho_\beta)=-S(\rho)-\Tr{\rho \ln \rho_\beta}\geq - S(\pi_\rho)-\Tr{\pi_\rho \ln \rho_\beta}=S(\pi_\rho||\rho_\beta)$, we get $\Delta F_{\beta}(\rho)\geq\Delta F_{\beta}(\pi_\rho)$.}.
In conclusion, our work provides an essential step forward for the field of quantum thermodynamics and we hope that it will inspire further investigations and applications of correlations in this field.

\end{document}